\documentclass{JACoW-GSI}
\usepackage{graphicx}
\usepackage{url}

\newcommand{\ee}{e^{+} e^{-}}

\newcommand{\leplep}{\ell^{+}\ell^{-}}
\newcommand{\jp}{J/\psi}
\newcommand{\psip}{\psi '}
\newcommand{\jpsi}{J/\psi}

\newcommand{\mumu}{\mu^{+}\mu^{-}}
\newcommand{\pipi}{\pi^{+}\pi^{-}}

\newcommand{\rt}{\rightarrow}

\begin{document}
\title{An Experimental Overview of the {\it X}, {\it Y} \& {\it Z} Charmoniumlike Mesons }

\author[]{Stephen L. Olsen\thanks{solsen@hep1.snu.ac.kr}}
%\author[1]{K. Musterfrau}
%\author[2]{C. Mustermann\thanks{cee@aps.anl.gov}}
\affil[]{Seoul National University, Seoul KOREA}
%\affil[2]{ANL, Argonne, IL 60439, USA}

\maketitle

{\bfseries A review of some of the recent experimental developments concerning 
the $X$, $Y$ and $Z$ charmoniumlike mesons states is presented.}

\section{Introduction}

The $X$, $Y$ \& $Z$ particles are an assortment of meson-resonance-like peaks that were discovered
by the BaBar and Belle $B$-factory experiments.  A common feature is that they are seen to decay
to final states that contain charmed ($c$) and anticharmed ($\bar{c}$) quarks and, thus, almost
certainly contain a $c\bar{c}$ quark pair among their constituent particles.  
The spectrum of conventional mesons that are comprised
of only a $c\bar{c}$ quark pair, {\it i.e.} the ``charmonium mesons,'' 
is generally considered to be the most well
understood hadronic system, both experimentally and theoretically, and most of the $XYZ$ candidate 
states do not match well to any of the remaining unassigned charmonium levels.  As a result, at 
least some of these states have been touted as candidates for ``exotic'' mesons, {\it i.e.}
mesons with a more complex substructure than the simple quark-antiquark anzatz of the venerable
Quark-Parton-Model (QPM).   

In particular, if the $Z$ states, seen by Belle as peaks in the 
$\pi^{+}\psi'$ and $\pi^{+}\chi_{c1}$ invariant mass distributions~\cite{conj} in 
$B\rightarrow K \pi^{+}\psi'$~\cite{belle_z4430} and 
$B\rightarrow K\pi^{+}\chi_{c1}$~\cite{belle_z14050}, respectively, are mesons, they
would necessarily have a minimal quark substructure of $c\bar{c}u\bar{d}$ and be, therefore,
manifestly exotic.  Here the experimental situation remains a bit uncertain in that an analysis
by the BaBar group does not confirm (or contradict) Belle's claim for the 
$Z(4430)^+\rightarrow \pi^{+}\psi'$ mass peak~\cite{babar_z4430}.  The
situation concerning the charged $Z$ states are discussed at this meeting by Ruslan Chistov (Belle),
Claudia Patrigiani (BaBar) and in a panel discussion chaired by Ryan Mitchell. 
I provide some of my own comments on the $Z$ states below.

Other topics covered here include: new results from Belle and CDF on the mass of the $X(3872)$;
a comment on the $J^{PC}$ determination of the $X(3872)$;  some
discussion on the $X$ and $Y$ states with masses near 3940~MeV including the first public
presentation of a new Belle study of the process $\gamma\gamma\rightarrow\omega J/\psi$,
which is dominated by a narrow peak near 3915~MeV.

\section{The states with mass near 3940~MeV}
In 2005, Belle reported observations of three states with masses near 3940~MeV: the $X(3940)$, seen
as a $D^*\bar{D}$ mass peak in exclusive $e^+e^-\rightarrow J/\psi D^*\bar{D}$ 
annihilations~\cite{belle_x3940}; the $Y(3940)$, seen as a near-threshold
$\omega J/\psi$ mass peak in the decay $B\rightarrow K\omega J/\psi$~\cite{belle_y3940};
and the $Z(3930)$, seen as a $D\bar{D}$ mass peak in untagged $\gamma\gamma\rightarrow D\bar{D}$
events~\cite{belle_z3930}.  Of these, only the $Z(3930)$ has been convincingly assigned to
a previously unfilled charmonium level. 

\paragraph{~} The $Z(3930)$ production angle distribution matches well the
$\sin^4\theta^*$ behavior expected for a $J=2$ meson and its mass ($3929\pm5\pm2$~MeV),
width ($29\pm 10\pm 2$~MeV) \& $\gamma\gamma$ production rate match well to expectations
for the $2^3P_2$ $c\bar{c}$ charmonium state, which is commonly called the $\chi_{c2}^{\prime}.$
There is general agreement that the $Z(3930)$ is, in fact, the $\chi_{c2}^{\prime}$.    

\paragraph{~}The $X(3940)$ is produced in association with a $J/\psi$ in the $e^+e^-\rt J/\psi X(3940)$
annihilation process, which unambiguously fixes its $C$-parity as $C=+1$.  Furthermore,
the only known charmonium states that are seen to be produced via the process 
$e^+e^-\rightarrow J/\psi (c\bar{c})$ have $J=0$, which provides some circumstantial evidence
that the $X(3940)$ has $J=0$.  This, taken together with the fact that  
the $X(3940)$ was discovered via its $D^*\bar{D}$ decay channel and is not seen 
to decay to $D\bar{D}$ -- a decay channel that is preferred for $0^{++}$ and
forbidden for $0^{-+}$ -- indicates that  
$J^{PC}=0^{-+}$ is its most likely quantum number assignment.  The unfilled $0^{-+}$
state with the closest expected mass value is the $3^1S_0$ $\eta_c^{\prime\prime}$, which
potential model predictions put at 4043~MeV (or higher)~\cite{barnes_prd72},
well above the $X(3940)$'s measured mass: $3942\pm 2\pm 6$~MeV~\cite{belle_prl100202001}.

\paragraph{~}The $Y(3940)$ mass is well above open-charm mass thresholds for decays
to $D\bar{D}$ or $D^*\bar{D}$ finally states, but was discovered via its decay to
the hidden charm $\omega\jp$ final state.  This implies an  $\omega\jp$ partial width
that is much larger than expectations for charmonium. 

\subsection{Are $X(3940)$ and $Y(3940)$ the same state?}
In a recently reported study of $B\rt KD^*\bar{D}$ decays, Belle searched
for, and did not find, a signal for $B\rt K Y(3940)$; $Y(3940)\rt D^*\bar{D}$~\cite{belle_0810-0358}.
The quoted upper limit on this mode corresponds to a lower limit on the 
branching fraction ratio:
\begin{equation}\label{eq:y3940_2_ddstr}
\frac{{\cal B}(Y(3940)\rt \omega\jp)} {{\cal B}(Y(3940)\rt D^{*0}\bar{D^0})}>0.75
\end{equation}
at the 90\% confidence level.
Likewise, Belle searched for evidence for $X(3940)\rt\omega\jp$ by searching for 
$\omega \jp$ systems recoiling from a $\jp$ in $\ee\rt \omega 2\jp$ annihilations~\cite{belle_x3940}.
Here no signal is seen and an upper limit
\begin{equation}\label{eq:x3940_2_wj}
\frac{{\cal B}(X(3940)\rt \omega\jp)} {{\cal B}(X(3940)\rt D^{*0}\bar{D^0})}<0.60
\end{equation}
was established at the 90\% CL.  These limits would be contradictory if the $X(3940)$ and
the $Y(3940)$ were the same state seen in different production modes.  Thus, the best current
evidence indicates that these two states are distinct.

\subsection{BaBar's confirmation of the $Y(3940)$}
In 2008, BaBar~\cite{babar_y3940} reported a study of $B\rt K\omega \jp$ 
in which the $\omega\jp$ invariant
mass distribution shows a near-threshold peaking that is qualitatively similar to
$Y(3940)$ peak previously reported by Belle.  
However, the BaBar values for mass and
width derived from fitting their data are both lower than the corresponding values reported
by Belle:  $M=3914^{+3.8}_{-3.4}\pm1.6$~MeV (BaBar) compared to
$3943\pm11\pm13$~MeV (Belle), and $\Gamma=33^{+12}_{-8}\pm0.6$~MeV (BaBar) compared
to $87\pm22\pm26$~MeV (Belle).  Part of the difference might be attributable to the
larger data sample used by BaBar (350~fb$^{-1}$ compared to Belle's 253 fb$^{-1}$),
which enabled them to use smaller $\omega\jp$ mass bins in their analysis. 

\subsection{Belle's new $\omega\jp$ mass peak in $\gamma\gamma\rt\omega\jp$}
New to this meeting is a report from Belle of a dramatic and rather narrow
peak in the cross section for $\gamma\gamma\rt\omega\jp$~\cite{belle_y3915}
that is consistent
with the mass and width reported for the $Y(3940)$ by the BaBar group.

\begin{figure}[htb]
\centering
\includegraphics*[width=75mm]{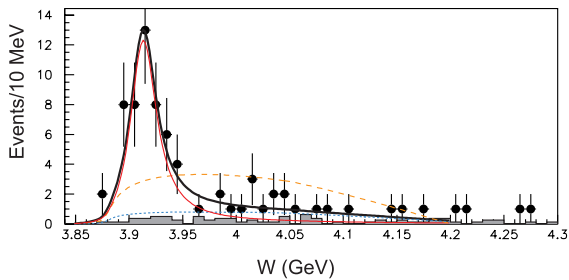}
\caption{The $\omega\jp$ mass distribution for selected events.}
\label{fig:omegajpsi}
\end{figure}

Belle selects events with $\pipi\pi^0$ and $\leplep$ ($\ell = \mu$~or~$e$)
tracks that have a net transverse momentum that less than 100~MeV.  In
events with $M_{\leplep}$ near $m_{\jp}$, the three pion system is found to 
be dominated by $\omega\rt\pipi\pi^0$ decays; likewise, in events where
$M_{3\pi}$ is near $m_{\omega}$, the dileptons are almost all from 
$\jp\rt\leplep$ decays.  After application of the requirements 
$|M_{3\pi}-m_{\omega}|<30$~MeV
\& $|M_{\leplep}-m_{\jp}|<25$~MeV and vetoing events with a 
$\psi'\rt\pipi\jp$, the invariant mass distribution for the
$\omega$ $\jp$ candidates, shown in Fig.~\ref{fig:omegajpsi},
shows a sharp peak near threshold and not much else.

The solid curve in Fig.~\ref{fig:omegajpsi} shows the result of a fit that
uses a phase-space-weighted, resolution-broadened
$S$-wave Breit Wigner (BW) function plus a smooth background function
that is forced to zero for masses below threshold.  The fit, which has a
$\chi^2/ndf=33.1/29$, gives {\em preliminary} results for the resonance parameters
of this peak, dubbed the $X(3915)$, of: 
\begin{eqnarray}\label{eq:omegajpsfit}
M & = & 3914\pm4\pm2~{\rm MeV};\\
\Gamma & = & 28\pm12 ^{+2}_{-8}~{\rm MeV};\\
N_{evts} & = & 60\pm 13 ^{+3}_{-14}.
\end{eqnarray}
The dashed curve in Fig.~\ref{fig:omegajpsi} shows the result of a fit with no
BW term.  The statistical significance of the signal, determined from the
square root of the change in likelihood for the fits with and without
a BW term and with the change in $ndf$
taken into account, is $7.1\sigma$.  The systematic errors on these parameters
are determined by varying the selection requirements and fitting procedure.

This preliminary value for the mass is about $2\sigma$ different from that of the
$Z(3930)$ ($M=3929\pm5\pm2$~MeV, indicating that these two peaks are distinct
and not different decay channels of the same state.  On the other hand, there
is good agreement between these preliminary results and the mass and width
quoted by BaBar for the ``$Y(3940)$,'' which is also seen in $\omega\jp$.
 
The $\omega\jp$ acceptance depends on the $J^{P}$ value. For $J^{P}=0^{+}$,
Belle determines
\begin{equation}\label{eq:3gammabr}
\Gamma_{\gamma\gamma}(X(3915)){\cal B}(X(3915)\rt\omega\jp)
=69\pm16^{+7}_{-18}{\rm eV},
\end{equation}
where $X(3915)$ is used to denote this new candidate state.

Whether or not the $X(3915)$ is the same as the $Y(3940)$, it 
has the same difficulty with a charmonium assignment. 
Using the total width measurement given above, Eq.~\ref{eq:3gammabr} can
be rewritten as:
$\Gamma_{\gamma\gamma}(X(3915))\Gamma(X(3915)\rt\omega\jp)\simeq 2000$~keV$^2$,
(albeit with large ($\sim \pm 50\%$) errors).  If for $\Gamma_{\gamma\gamma}$ we
apply a value that is typical for charmonium, {\i.e. $1\sim 2$~keV), we find a partial
$\Gamma(X(3915)\rt\omega\jp)\sim {\mathcal O}$(1~MeV), which is quite large for charmonium.
Here a $J^P = 2^+$ assignment would help some, but not too much. 

\section{The $X(3872)$}

The $X(3872)$ was discovered by Belle in 2003~\cite{belle_x3872} as a narrow
peak in the $\pipi\jp$ invariant mass distribution from $B^+\rt K^+\pipi\jp$ decays.
This peak was subsequently confirmed by
CDF~\cite{CDF_x3872}, D0~\cite{D0_x3872} and BaBar~\cite{babar_x3872}.
CDF and D0 see $X(3872)$ produced promptly in inclusive $p\bar{p}$ collisions as well
as in $B$ meson decays.   In all of the experiments, the invariant mass distribution of the
dipion system is consistent with originating from $\rho\rt\pipi$~\cite{CDF_pipi}.  
If this is the case, the
$C$-parity of the $X(3872)$ must be $C=+1$.  Charmonium states are all Isosinglets;
the decay charmonium$\rt\rho\jp$ violates Isospin and should be strongly suppressed.

\subsection{Comment on the $J^{PC}$ value of the $X(3872)$}
A study of angular correlations among the $\pipi\jp$ final state particles by CDF led them to conclude
that the only likely $J^{PC}$ assignments for the $X(3872)$ are $1^{++}$ and $2^{-+}$, with $1^{++}$ 
preferred~\cite{CDF_jpc}.  Subsequently, the $2^{-+}$ assignment has been further
disfavored by BaBar's report of $>3\sigma$ significance signals for $X(3872)$ decays
to both $\gamma\jp$ and $\gamma\psi'$~\cite{babar_gammajpsi}.  The radiative transition of a
$2^{-+}$ state to the $\jp$ or $\psi'$ would have to proceed via a higher order
multipole term and be highly suppressed.   For these reasons, the most likely $J^{PC}$
is $1^{++}$.

\subsection{The $X(3872)$ mass}
An intriguing feature of the $X(3872)$ is its close proximity in mass to
the $D^{*0}\bar{D^0}$ mass threshold.  This has stimulated a number
of papers that interpret the $X(3872)$ as a molecule-like arrangement comprised of
a $D^{*0}$- and $\bar{D^0}$-meson~\cite{molecule}. 
Critical to these models is whether the $X(3872)$  mass is above
or below $m_{D^{*0}}+m_{D^0}$.    
In 2008, Belle reported a new result for the mass of the $X(3872)$ determined using
the  $X(3872)\rt\pipi\jp$ decay mode:
$M^{Belle}_{X(3872)}=3871.46\pm0.37\pm0.07$~MeV~\cite{belle_x3872_mass}. 
This year, the CDF group reported an even more precise measurement of the mass using
the same decay channel:  $M^{CDF}_{X(3872)}=3871.61\pm0.16\pm0.19$~MeV~\cite{CDF_x3872_mass}.  
A new world average that includes
these new measurements plus other results that use the $\pipi\jp$ decay mode
is $M^{avg}_{X(3872)}=3871.46\pm0.19$~MeV. This puts
the $X(3872)$ within about one part in $10^{4}$ of the $D^{*0}\bar{D^0}$ mass threshold:
$m_{D^{*0}}+m_{D^0} = 3871.81 \pm 0.36$~MeV~\cite{PDG}, and sets the binding energy of any
possible $D^{*0}\bar{D^0}$ component of the $X(3872)$ at $-0.35\pm0.41$~MeV.
Note that any significant improvements in the precision of this quantity will
require improvement in the $D^0$ mass determination, which is currently known to within
$\pm180$~keV~\cite{PDG}.  This is something that BES-III could provide.

\subsection{Are there $X(3872)$ partner states?}
Another interpretation suggests that the $X(3872)$ is a tightly bound diquark-diantiquark
system~\cite{maiani_1,ebert}.  In this picture the existence of nearby partner states is expected.
The observed $X(3872)$, which is produced in $B^+$ decays, is interpreted as a $cu\bar{c}\bar{u}$
combination (dubbed $X_L$).  In $B^0\rt K_S\pipi\jp$, one should see
a partner state, the $X_h = cd\bar{c}\bar{d}$ combination,which
differs in mass by $8\pm3$~MeV~\cite{mass-diff}.
In addition, Isospin and Flavor-$SU(3)$ partner states
({\it e.g.,} $X^+ = cu\bar{c}\bar{d}$  and $X_s =cs\bar{c}\bar{d}$) are also expected to exist. 

BaBar searched for a charged version of the $X(3872)$ in the $\pi^-\pi^0\jp$ mass distribution
in $B \rt K\pi^-\pi^0\jp$ decays and found no evidence for a signal in either $B^0$ or $B^+$
decays~\cite{babar_xplus_prd71031501}. The BaBar 90\% CL upper limit on the number of
$B^0\rt K^+ X^-$ events is 15.9 events, which should be
compared to the Isospin symmetry expectation of $75\pm25$.  They rule out an isovector
hypothesis for the $X(3872)$ with 99.99\% confidence.

Both Belle~\cite{belle_x3872_mass} and BaBar~\cite{babar_deltam} measured the $X(3872)$ mass for 
$B^+\rt K^+\pipi\jp$ and $B^0\rt K_s\pipi\jp$ decays separately.  They both find mass differences
that are consistent with zero:  $M_{X_H} - M_{X_L} = 0.2\pm0.9\pm0.3$~MeV for Belle and
$2.7\pm1.6\pm0.4$~MeV for BaBar.  The CDF group tried fitting their $\sim$6000 event
$X(3872)\rt \pipi\jp$ peak with two different mass Gaussians, they rule out a mass difference
of less that 3.6~MeV (95\% CL) for equal $X_H$ and $X_L$ production~\cite{CDF_x3872_mass}.

\subsection{$X(3872)\rt D^{*0}\bar{D^0}$}

With a data sample containing 447M $B\bar{B}$ meson pairs,
Belle observed a near-threshold $D^0\bar{D^0}\pi^0$ mass enhancement 
in $B\rt K D^0\bar{D^0}\pi^0$ decays that, when interpreted as 
$X(3872)\rt D^0\bar{D^0}\pi^0$, gave an $X(3872)$ mass of 
$3875.4\pm 0.7 ^{+1.2}_{2.0}$~MeV~\cite{belle_x3872_ddpi}.
BaBar studied $B\rt KD^{*0}\bar{D^0}$ with a sample of 383M $B\bar{B}$ pairs and found
a similar near-threshold enhancement that, if considered to be due the the $X(3872)\rt D^{*0}\bar{D^0}$,
gave a mass of $3875.1^{+0.7}_{0.5}\pm0.5$~MeV~\cite{babar_x3872_ddstr}. 
These mass values are distinctly higher than that
seen for the $\pipi\jp$ channel and this raised some hope that these may be the neutral partner
state predicted by the diquark-diantiquark model.  However, a subsequent Belle study of 
$B\rt KD^{*0}\bar{D^0}$ based on 657M $B\bar{B}$ pairs finds a mass for the near threshold peak of 
$3872.9 ^{+0.6} _{-0.4}$ $^{+0.4} _{-0.5}$~MeV, much closer to the value determined from the $\pipi\jp$ 
decay channel.

In the meantime, Braaten and co-authors have pointed out that in a 
narrow decaying $D^{*0}\bar{D^0}$ molecular system
the decays of the constituent $D^{*0}$ are important and the width of the $D^{*0}$ distorts
the decay line shape in this channel~\cite{braaten_0709.2697,braaten_0907.3167}.  
Therefore, fitting the $D\bar{D}\pi$ or $D^*\bar{D}$ to a BW function, as the experiments have done,
does not give reliable values for either the mass or width.

\subsection{Belle study of $B\rt K\pi X(3872)$}
If, in fact, the $X(3872)$ is a $D^{*0}\bar{D^0}$ molecule, it is a very strange object.
The small value for the binding energy given above means that the constituent
$D^{*0}$ and $\bar{D^0}$ are generally very far apart in space:  for the central
value, {\it i.e.} $E_B = 0.25$~MeV, their rms separation would be a huge 
6~fermis or higher~\cite{braaten_0907.3167}.
In such a case, the constituent $D^*$ and the $\bar{D}$ would rarely be near enough to
each other to allow for the formation of a $\jp$, which has to happen for the
$\pipi\jp$ decay to occur. Likewise, it would seem that the prompt production of such a 
fragile object in high energy $p\bar{p}$ collisions, as seen by CDF~\cite{CDF_x3872} and
%D.~Acosta etal (CDF) PRL 93 072001 (2004)
D0~\cite{D0_x3872}, would also be improbable.  In fact, the production characteristics
%V.M.~Abazov et al (D0), PRL 93,162002(2004)
of the $X(3872)$ in $\sqrt{s}=1.96$~GeV $p\bar{p}$ collisions, such as the $p_T$ \&
rapidity distributions and the ratio of prompt production {\it vs.} production 
via $B$-meson decays, are very similar to those of the 
well established $\psip$ charmonium state~\cite{D0_x3872,CDF_x3872_prompt-ratio}.
%G.~Bauer, Int. J. Mod. Phys. A20, 3767 (2005)
 
To get around this, molecule advocates usually conjecture that the
physical $X(3872)$ is a quantum mechanical mixture of a $D^*\bar{D}$ molecule
and the $2^3P_1$ $c\bar{c}$ charmonium state ({\it i.e.} the $\chi^{\prime}_{c1}$) and the
latter component dominates the production and decays to final states that contain charmonium. 
Therefore it is of interest to compare production characteristics of the $X(3872)$ to those
of other charmonium states in $B$-meson decays.  One common characteristic of all of 
the known charmonium states that are produced in $B$ meson decays is that when 
they are produced in association with a $K\pi$ pair, the $K\pi$ system is always
dominated by a strong $K^*(890)\rt K\pi$ signal.  

Belle did a study of $X(3872)$ production in association with a $K\pi$ in 
$B^0\rt K^+\pi^- \pipi\jp$ decays~\cite{belle_x3872_mass}.  In a sample of 657M $B\bar{B}$ pairs
they see a signal of about 90 events where the $\pipi\jp$ comes from
$X(3872)$ decay.  Figure~\ref{fig:kpix3872} shows the $K\pi$ invariant
mass distribution for these events, where it is evident that most of the
$K\pi$ pairs have a phase space-like distribution, with little or
no signal for $K^*(890)\rt K\pi$.  This should be contrasted to the $B\rt K\pi \psi'$ 
events (with $\psi'\rt \pipi\jp$) events in the same data sample,  where the
$K\pi$ invariant mass distribution, shown in Fig.~\ref{fig:kpipsip}, is dominated
by the $K^*(890)$.  

\begin{figure}[htb]
\centering
\includegraphics*[width=75mm]{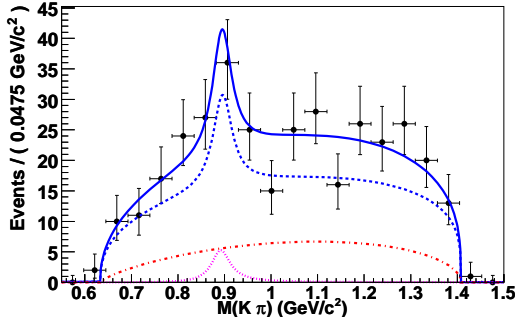}
\caption{The $K\pi$ mass distribution for $B\rt K\pi X(3872)$ events
from ref.~\cite{belle_x3872_mass}.}
\label{fig:kpix3872}
\end{figure}

\begin{figure}[htb]
\centering
\includegraphics*[width=75mm]{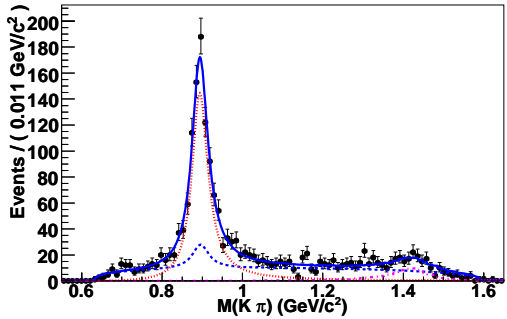}
\caption{The $K\pi$ mass distribution for $B\rt K\pi \psi'$ events from ref.~\cite{belle_x3872_mass}.}
\label{fig:kpipsip}
\end{figure}

Belle reports a $K^*(890)$ to $K\pi$ non-resonant ratio of 
\begin{equation}
\frac{{\mathcal B}(B\rt (K^+\pi^-)_{K^*(890)}\jp)}{{\mathcal B}(B\rt (K^+\pi^-)_{NR}\jp)}<0.55,
\end{equation}
For comparison, from branching fractions listed in the PDG, I
estimate the corresponding ratio for $B\rt K^+\pi^-\jp$ decays to be $\sim$3.0, albeit with a large
error.    

\section{The $1^{--}$ states produced by ISR}

Thanks to the very high luminosities enjoyed by the $B$-factory experiments, while they run
at the $\Upsilon(4S)$ ($\sqrt{s}=10.58$~GeV) and nearby continuum, they also accumulate 
lots of $\ee$ annihilation
data at lower energies via the initial-state-radiation process $\ee\rt\gamma_{ISR} X$.  When
the ISR gamma-ray energy is in the $4\sim 5$~GeV range, the $\ee$ annihilation occurs in the
$\sqrt{s'}=3\sim 5$~GeV range, the energy region populated by charmonium states.   The BaBar group
used the ISR process to study the cross section for $\ee\rt\pipi\jp$ in the charmonium region
and discovered a large, relatively broad peak near $4260$~MeV~\cite{babar_y4260}. 
BaBar's fitted mass for this peak, which they call the $Y(4260)$, is $M=4259\pm 8^{+2}_{-6}$~MeV and 
its total width is $\Gamma=88\pm 23^{+6}_{-4}$~MeV.  The $Y(4260)$ was confirmed by both
CLEO~\cite{CLEO_y4260} and Belle~\cite{belle_y4260}.  Belle cross-section measurements
for $\ee\rt \pipi\jp$ in the $\sqrt{s}= 4 \sim 5$~GeV 
region are shown in Fig.~\ref{fig:belle_y4260}, where the cross section
at the $Y(4260)$ peak is $\sim 70$~pb.

\begin{figure}[htb]
\centering
\includegraphics*[width=75mm]{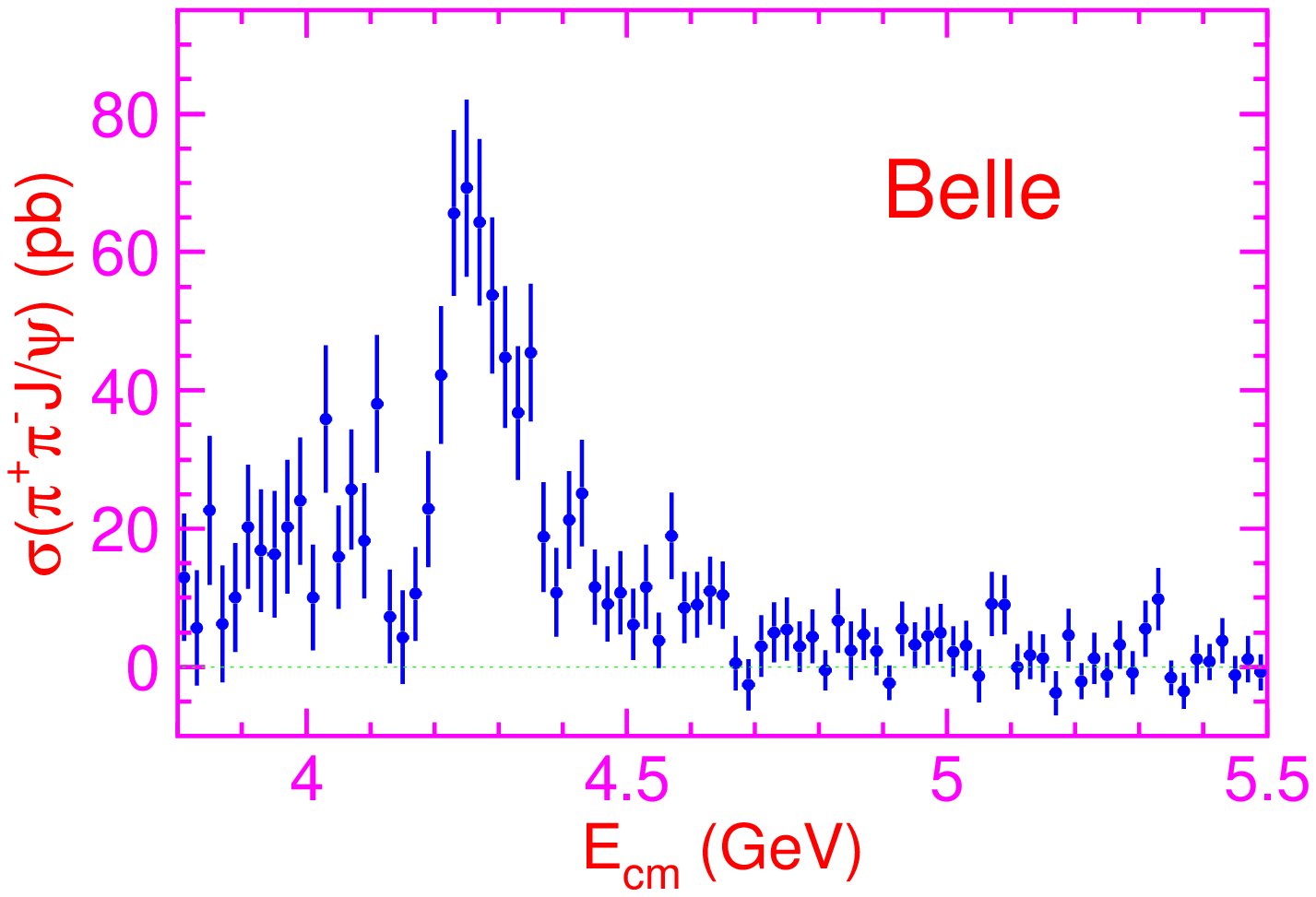}
\caption{Cross sections for $\ee\rt \pipi\jp$.}
\label{fig:belle_y4260}
\end{figure}

\begin{figure}[htb]
\centering
\includegraphics*[width=75mm]{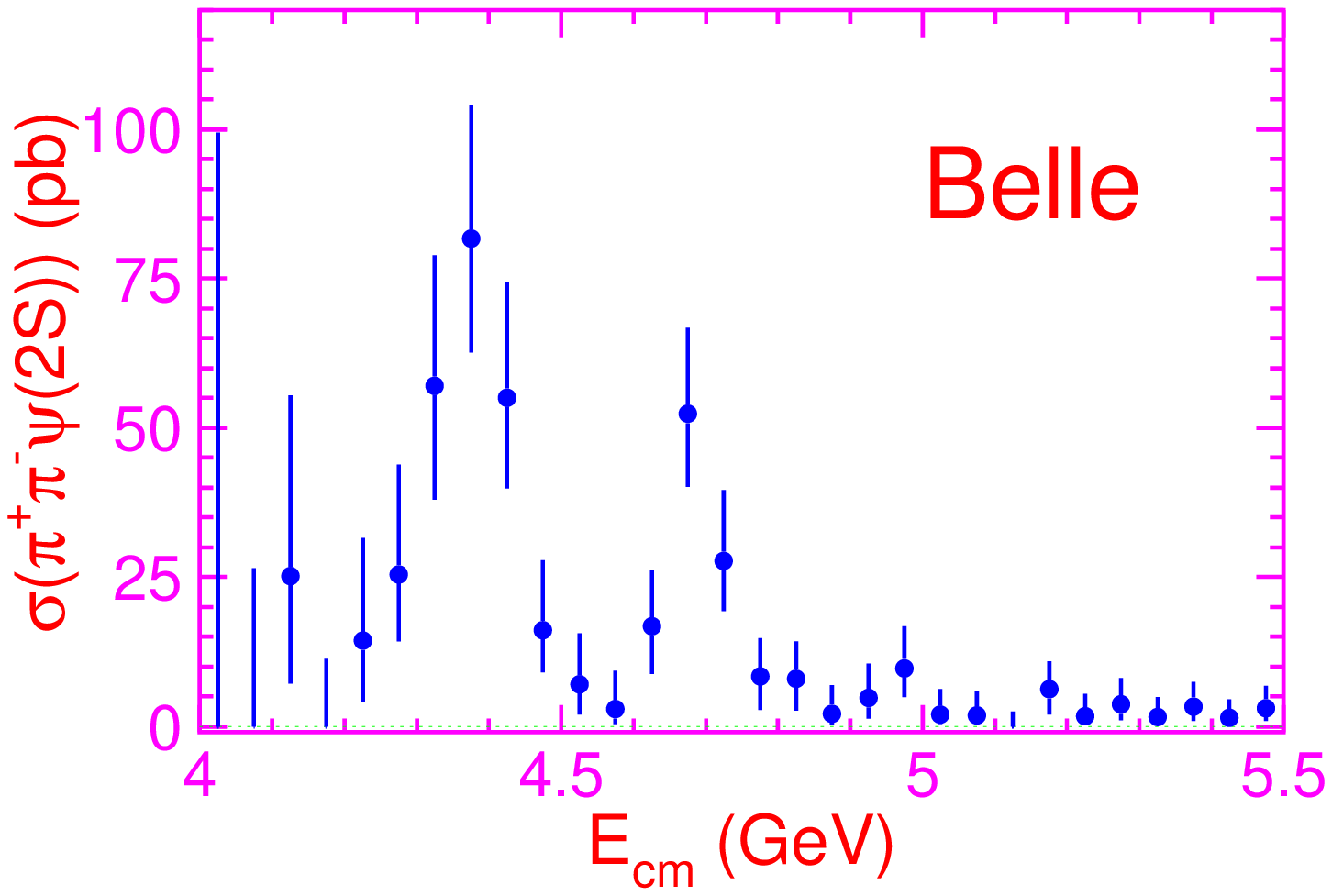}
\caption{Cross sections for $\ee\rt \pipi\psip$.}
\label{fig:belle_y4325}
\end{figure}

The BaBar group subsequently reported
a similar structure in the cross section for $\ee\rt \pipi\psi'$, but
in this case the fitted mass, $M=4324\pm 24$~MeV and width $\Gamma=172\pm 33$~MeV are
both significantly higher than the values found for the $Y(4260)$~\cite{babar_y4325}.  
Belle confirmed the general features of the BaBar $\pipi\psi'$ peak but,
thanks to a larger data sample (673 fb$^{-1}$ for Belle compared to 272~fb$^{-1}$ for BaBar) they
found that the structure is formed from two narrower peaks.  
Belle's fit to these two peaks give
$M_1=4361\pm 9\pm 9$~MeV \& width $\Gamma_1=74\pm 15\pm 10$~MeV (the $Y(4360)$)
$M_2=4664\pm 11\pm 5$~MeV \& $\Gamma_2=48\pm 15\pm 3$~MeV (the $Y(4660)$)~\cite{belle_y4325}. 
Figure~\ref{fig:belle_y4325}
shows Belle's $\ee\rt\pipi\psip$ cross section measurements, where the two peak values corresponding 
to the $Y(4369)$ and the $Y(4660)$ are  $\sim 80$~pb \&
$\sim 50$~pb, respectively, and similar to the peak cross-section
value for the $Y(4260)$ shown in Fig.~\ref{fig:belle_y4260}.

\subsection{Can these be charmonium states?}

There is only one unassigned $1^{--}$ charmonium state in this mass region, the $3^3D_1$ level.  This
might accommodate the $Y(4660)$, but there is no room in the spectrum for all three of the peaks
discussed above. 
A tantalizing feature of all three of these states is the total absence of any corresponding peaking
features in the total cross section for $\ee$ annihilation into
hadrons at the same energy.  Figure~\ref{fig:bes-Rhad} shows BES measurements of 
$R_{had}=\sigma(\ee\rt {\rm hadrons})/\sigma_{QED}(\ee\rt \mumu)$ in the same energy region,
where the cross section exhibits dips near the locations of the 
$Y(4260)$ and $Y(4360)$~\cite{bes_Rhad}.   (The
BES $R_{had}$ measurements do not span the $Y(4660)$ region.)

\begin{figure}[htb]
\centering
\includegraphics*[width=65mm]{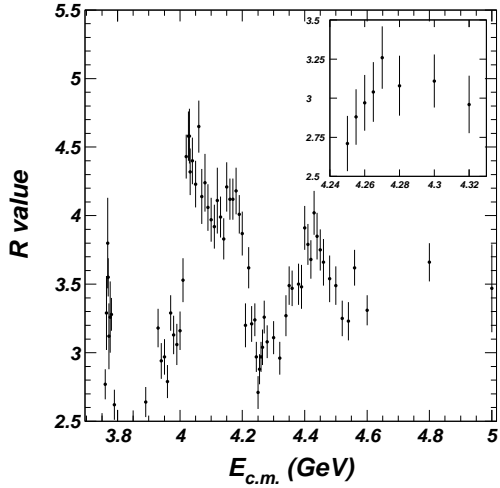}
\caption{The cross section for $\ee\rt$~hadrons in the charmonium region measured by
BES (from ref.~\cite{moxh}).}
\label{fig:bes-Rhad}
\end{figure}

The absence of any evidence for $Y(4260)$ ($Y(4360)$) decays to open charm implies that the 
$\pipi\jp$ ($\pipi\psip$) partial width is large: the analysis of ref.~\cite{moxh} 
gives a 90\% CL lower limit $\Gamma(Y(4260)\rt\pipi\jp)>508$~keV, which should be compared to the
corresponding $\pipi\jp$ partial widths 
of established $1^{--}$ charmonium states:
89.1~keV for the $\psi'$ and 44.6~keV for the $\psi''$~\cite{PDG}.

Belle and BaBar have exploited ISR to make measurements of cross sections
for exclusive open-charm final states in this 
energy range~\cite{pakhlova_ddstr,pakhlova_lclcbar}.  These are 
discussed in detail at this meeting by Galina Pakhlova.  She reports
that the exclusive channels that have been measured so far 
--- the sum of which very nearly saturates the total inclusive cross section --- show 
no evidence for peaking near the masses of the $Y$ states. The one exception
is $\ee\rt\Lambda^+_c \Lambda_c^-$, which has a threshold peak in the vicinity
of the $Y(4660)$ peak mass~\cite{pakhlova_lclcbar}.

\subsection{Search for 
$Y(4260)\rt D^{(*)}\bar{D}\pi$ using ISR}

The most commonly invoked theoretical explanation for the ISR-produced
$1^{--}$ $Y$ states is that they are $c\bar{c}$-gluon hybrids~\cite{hybrids}, 
{\it i.e.} mesons containing a $c\bar{c}$ pair plus an excited gluonic field.  
From this point of view, the lack of any 
evidence for $D^{(*)}\bar{D}^{(*)}$ decays is explained by the 
theoretically motivated expectation that
the relevant open-charm thresholds for $c\bar{c}$-gluon hybrids 
are $M_{D^{**}} + M_{D}$, where $D^{**}$ designates the low-lying
$P$-wave charmed mesons: the lowest of these are the very wide $J^P = 0^+$
$D_0(2400)$ with $M\simeq 2350$~MeV and $\Gamma \simeq 260$~MeV,
and the narrow $J^P = 1^+$ $D_1(2420)$ with $M \simeq 2420$~MeV and 
$\Gamma \simeq 20$~MeV.  Note that there is considerable overlap
between the broad $Y(4260)$ peak and the thresholds
for both $D^{**}=D_0(2400)$ and $D^{**}=D_1(2420)$. 
The prominent decay modes of the $D_0(2400)$ 
and $D_1(2420)$ are $D\pi$ and $D^*\pi$, respectively.  Therefore,
searches for the $Y(4260)$ in {\it both} the exclusive $\ee\rt 
D\bar{D}\pi$ and $D^*\bar{D}\pi$ channels are especially important.

In 2008, Belle~\cite{pakhlova_ddbarpi}  published the ISR measurements of
$\sigma(\ee\rt D^0 D^-\pi)$ shown in Fig.~\ref{fig:y4260_ddbarpi},
show a strong $\psi(4415)$ signal. (This is seen to be
due to $\psi(4415)\rt D^*_2(2460)\bar{D}$, where $D_2^*(2460)$ is the
$J=2$ $D^{**}$ state, and this observation
strongly supports the $\psi(4415)$ assignment to the
$\psi(4S)$ charmonium state~\cite{barnes_prd72}.)  
%T.~Barnes, S.~Godfrey and E.S.~Swanson PRD 72 054026 (2005)
However, the data show no indication of a
$Y(4260)\rt D_0(2400)\bar{D}$ signal as expected for
a $c\bar{c}$-gluon hybrid assignment for the $Y(4260).$
In fact, the cross section
is consistent with zero throughout the $Y(4260)$ mass region, at least within
the $\sim\pm 100$~pb errors of the data points.  Note that the
cross section (in Fig.~\ref{fig:belle_y4260} above) for $\ee\rt\pipi\jp$ 
at the $Y(4260)$ peak is $\sim 70$~pb, which
indicates that $Y(4260)\rt D_0(2400)\bar{D}$ decays cannot 
be much more frequent than $Y(4260)\rt \pipi\jp$ decays. 

\begin{figure}[htb]
\centering
\includegraphics*[width=75mm]{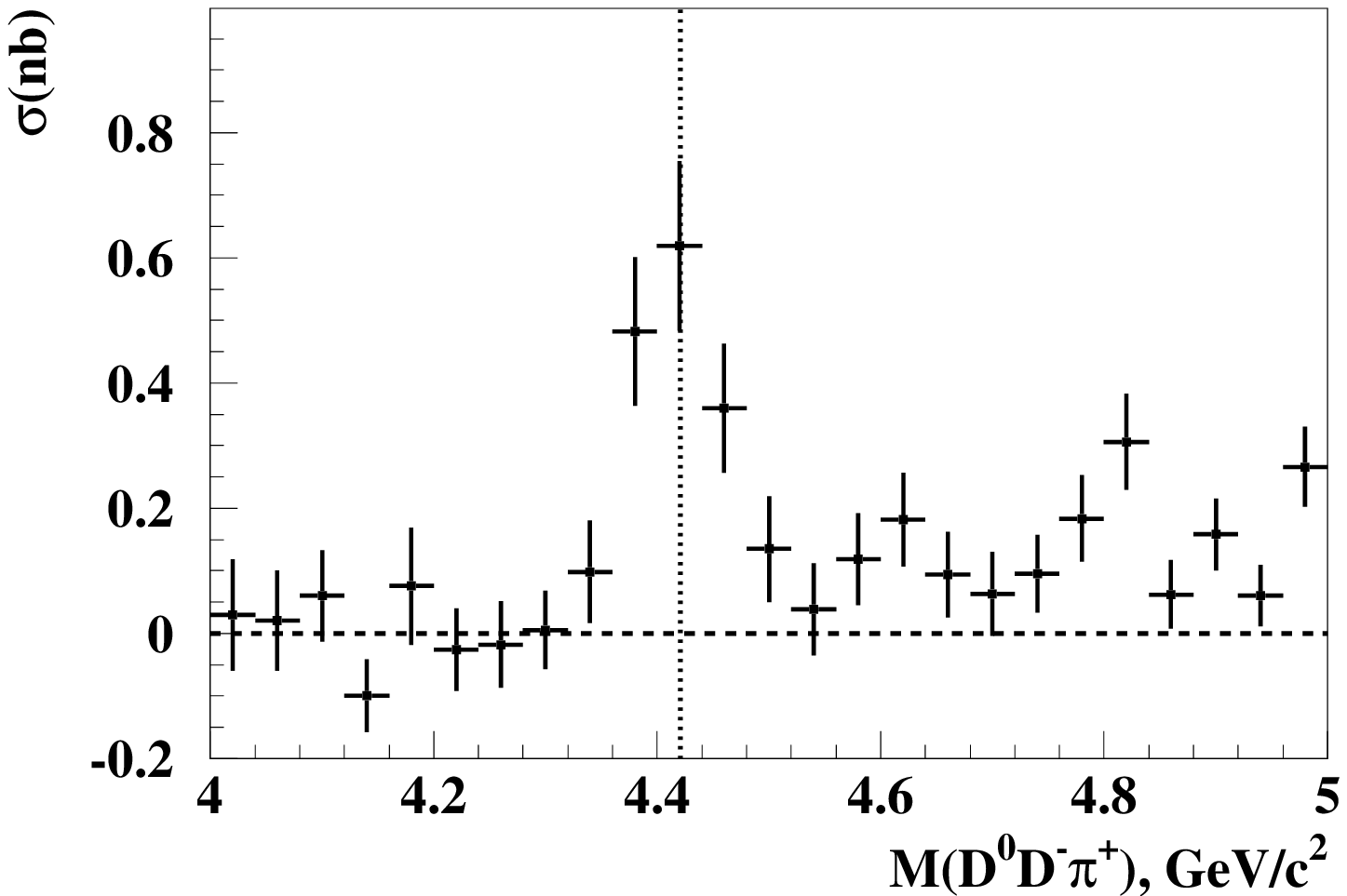}
\caption{ $\sigma(\ee \rt 
D^0 D^-\pi^+)$ from ref.~\cite{pakhlova_dstrdbarpi}.}
\label{fig:y4260_ddbarpi}
\end{figure}

In this meeting, 
Galina Pakhlova provided the first report of new
Belle results for $\sigma(\ee\rt D^{*-}\bar{D}^0 
\pi^+)$ shown in Fig.~\ref{fig:y4260_ddstrpi}~\cite{pakhlova_dstrdbarpi}. 
Here, although the error bars are larger,
there is also no sign at all of a $Y(4260)$ signal (or, for that matter, a $Y(4350)$ signal,
or a $Y(4660)$ signal).   The curve in the figure shows a fit that includes
a $\psi(4415)$ term and a smooth background;  the $\psi(4415)$ signal yield
from this fit is $14.4\pm 6.2 ^{+1.0}_{-9.5}$ events with a statistical
significance of $3.1\sigma$.

\begin{figure}[htb]
\centering
\includegraphics*[width=75mm]{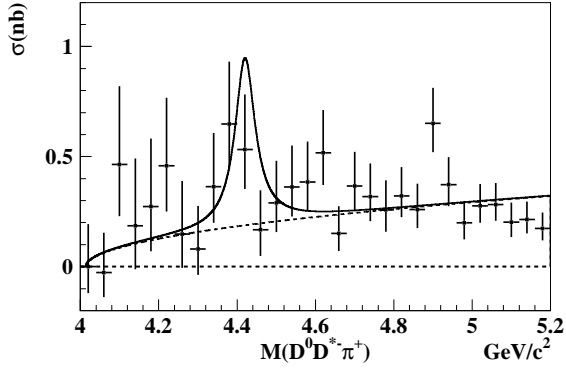}
\caption{ $\sigma(\ee \rt 
D^*\bar{D}\pi)$ distribution from ref.~\cite{pakhlova_dstrdbarpi}. 
The curve show results of the fit described in the text.}
\label{fig:y4260_ddstrpi}
\end{figure}

A fit to the data in Fig.~\ref{fig:y4260_ddstrpi} using two incoherent Breit-Wigner
functions, one to represent the $Y(4260)$ and the other for the $\psi(4415)$, plus an
incoherent smooth background term give a 90\% CL upper limit on
${\cal{B}}(Y(4260)\rt D^0 D^{*-}\pi^+)/
{\cal{B}}(Y(4260)\rt \pipi\jpsi)<15 $.  Consideration of the possibility of coherent
destructive interference between the different fit components could
inflate this upper limit by as much as a factor of four, but even this would be
pretty small compared to ratio between branching fractions for specific
open-charm  modes and that for $\pipi\jp$ for the $\psi(3770)$ charmonium state,
which are of the order $\sim 250$~\cite{PDG}.  Similar limits obtain for the $Y(4350)$
\& $Y(4660)$.  These results are discussed in Galina Pakhlova's
report in these proceedings.

\section{The charged $Z$ states}
 
Belle's $Z(4430)^+$ signal is the sharp peak  in  
the $\pi^+\psi'$ invariant mass distribution  from $B\rt K\pi^+\psi'$ decays
shown in Figure~\ref{fig:z4430_mpipsip}~\cite{belle_z4430}.   A fit using 
a BW resonance function gives a
mass of $M=4433\pm 4\pm 2$~MeV and total width of $\Gamma = 45^{+18~~+30}_{-13~~13}$~MeV,
with an estimated statistical significance of more than $6\sigma$.  Consistent signals
are seen in various subsets of the data: {\it i.e.} for both the $\psi'\rt \ell^+\ell^-$
\& $\psi'\rt\pipi\jp$ subsamples, the $\psi'$($\jp$)$\rt\ee$ \& $\mumu$ subsamples, etc.

\begin{figure}[htb]
\centering
\includegraphics*[width=65mm]{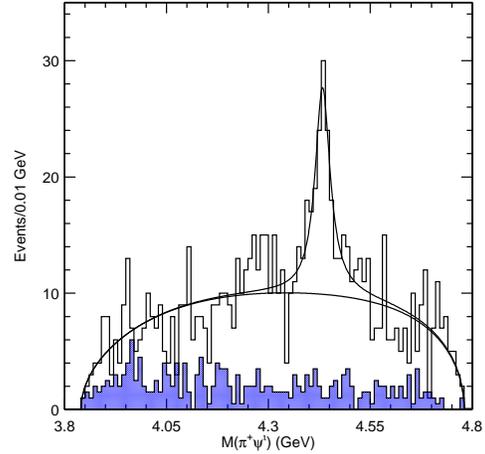}
\caption{The $\pi^+\psi'$ invariant mass distribution for $B\rt K\pi^+\psi'$ decays
(from ref.~\cite{belle_z4430}).}
\label{fig:z4430_mpipsip}
\end{figure}

Figure~\ref{fig:z4430_dalitz} shows the Dalitz plot for the $B\rt K\pi^+\psi'$ event candidates,
where vertical bands for $K^*(890)\rt K\pi$ and $K^*_2(1430)\rt K\pi$ are evident
and the $Z(4430)$ shows up as a horizontal band of events between $M^2(\pi\psi')=19~\&~20$~GeV$^2$.
(In the $M(\pi\psi')$ distribution of Fig.~\ref{fig:z4430_mpipsip}, the the $K^*$ bands are
suppressed by cuts on the $K\pi$ masses.)

\begin{figure}[htb]
\centering
\includegraphics*[width=65mm]{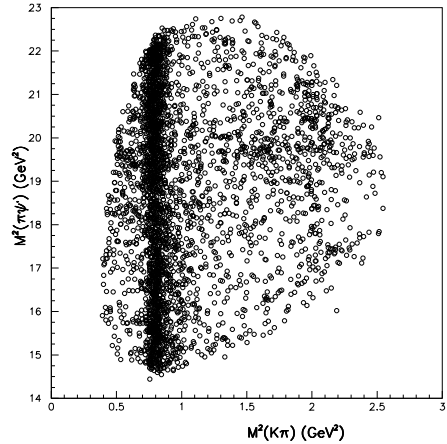}
\caption{The $M^2(K\pi)$ (horizontal) {\it vs.} $M^2(\pi\psi')$ (vertical) Dalitz plot distribution 
for candidate $B\rt K\pi\psi'$ events (from ref.~\cite{belle_z4430}).}
\label{fig:z4430_dalitz}
\end{figure}

\subsection{Is the $Z(4430)^+$ a reflection from $K\pi$ dynamics?}

A danger in searching for resonant structures in the $\pi\psi'$ channel in three-body $B\rt K\pi\psi'$
decays is the possibility that dynamics in the $K\pi$ channel can cause mass structures in
the $\pi\psi'$ invariant mass distribution that have no relation to $\pi\psi'$ dynamics.
This is because energy-momentum conservation imposes a tight correlation between
the decay angle ($\theta_{\pi}$) in the $K\pi$ system~\cite{theta-pi_def}
and the $\pi\psi'$ invariant mass. In fact,
$M^2(\pi\psi')$ is very nearly proportional to $\cos\theta_{\pi}$.  As a result, interference
between different partial waves in the $K\pi$ system can produce peaks in the $M(\pi\psi')$ 
that are merely ``reflections'' of structures in $\cos\theta_{\pi}$.  However, 
in the kinematically allowed $K\pi$ mass range for $\rt K\pi\psip$ decay,
only $S$, $P$ and $D$ $K\pi$ partial waves are significant,
and this limited set of partial waves 
can only produce fake $\pi\psip$ mass peaks at a discrete set of mass values.  

In the case of the $Z(4430)$, the $\pi^+\psip$ peak mass corresponds to $\cos\theta_{\pi}\simeq 0.25$,
and it is not possible to produce a peak near $\cos\theta_{\pi}\simeq 0.25$ 
with any combination of interfering $L=0,~1~\&~2$ partial waves without introducing larger additional
structures at other $\cos\theta_{\pi}$ values.  This is illustrated in Fig.~\ref{fig:z4430_coskpi},
where the histogram shows the distribution of $\cos\theta_{\pi}$ values for a MC sample of
$B\rt K Z(4430)$, $Z(4430)\rt\pi\psi'$ events where the $Z$ mass and width closely correspond to Belle's
reported values.  The curves in the figure show the results of trying to make a peak at at
the same location with interfering $S$, $P$ and $D$ partial waves in the $K\pi$ channel.  
(Here both longitudinally
and transversely polarized $\psi'$'s are considered, and no attempt is made
to restrict the strength of each term to that seen for the $S$-, $P$- and $D$-wave $K\pi$ components
in the data.)  These curves show that although a peak can be made at $\cos\theta_{\pi}\simeq 0.25$,
it is necessarily accompanied by much larger peaks near $\cos\theta_{\pi}\simeq\pm 1$.  No
such structures are evident in the $\pi\psi'$ mass plot of Fig.~\ref{fig:z4430_mpipsip}.

\begin{figure}[htb]
\centering
\includegraphics*[width=65mm]{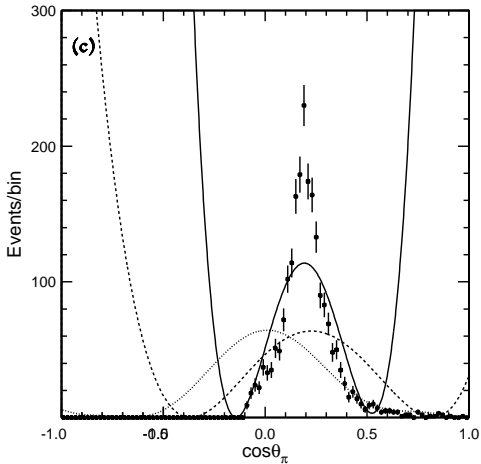}
\caption{The  histogram shows the $\cos\theta_{K\pi}$ distribution for a 
MC-generated $\pi\psi'$ resonance with $M=4.43$~GeV and $\Gamma = 0.05$~GeV.  
The curves show the results of attempts to produce a peak in the vicinity of
the data with interfering $S$, $P$ and $D$ waves in the $K\pi$ channel.}
\label{fig:z4430_coskpi}
\end{figure}
   
\subsection{A Dalitz analysis of $B\rt K\pi\psip$}

After the BaBar group did not confirm~\cite{babar-signif}
the $Z(4430)^+\rightarrow \pi^{+}\psi'$ mass peak
in their analysis of $B\rt K\pi\psip$ decays~\cite{babar_z4430}, the Belle group performed a 
reanalysis of their data that took detailed account of possible reflections from the $K\pi$
channel.  Specifically, they modeled the $B\rt K\pi\psip$ process as the sum of
two-body decays $B\rt K_i^*\psi'$, where $K_i^*$ denotes all of the known $K^*\rt K\pi$ resonances
that are kinematically accessible, and both with and without a $B\rt KZ$ component, where
$Z$ denotes a resonance that decays to $\pi\psi'$~\cite{belle_z4430_dalitz}.  The results of
this analysis, details of which are provided by Ruslan Chistov in these 
proceedings, confirm
the basic conclusions of Belle's 2007 publication.   

The data points in 
Fig.~\ref{fig:z4430_dalitz-analysis} shows the $M^2(\pi\psip)$ Dalitz plot projection with 
the prominent $K^*$ bands removed (as in Fig.~\ref{fig:z4430_mpipsip}) compared with the results
of the fit with no $Z$ resonance, shown as a dashed histogram, and that with a $Z$ resonance,
shown as the solid histogram.  The fit with the $Z$ is favored over the fit with no $Z$ by
$6.4\sigma$.  The fitted mass, $M=4443^{+15~~+19}_{-12~~-13}$~MeV, agrees within the systematic
errors with the earlier Belle result;  the fitted width, 
$\Gamma = 107^{+86~~+74}_{-43~~-56}$~MeV, is  larger, but also within the new 
analysis'systematic errors of the previous result.  
In the default fit, the $Z$ resonance was assumed to have zero spin.
Variations of the fit the included a $J=1$ assignment for the $Z$
as well as models that included additional, hypothetical $K^* \rt K\pi$
resonances with floating masses and widths, and radically different parameterizations
of the $K\pi$ $S$-wave amplitude do not change the conclusions~\cite{model_variations}
The product branching fraction from the Dalitz fit: 
${\mathcal B}(B^0\rt K Z^+)\times {\mathcal B}(Z^+ \rt\pi^+\psip) 
= (3.2^{+1.8~+9.6}_{-0.9~-1.6})\times 10^{-5}$ is not in strong contradiction with the
BaBar 95\% CL upper limit of $3.1\times 10^{-5}$.

\begin{figure}[htb]
\centering
\includegraphics*[width=65mm]{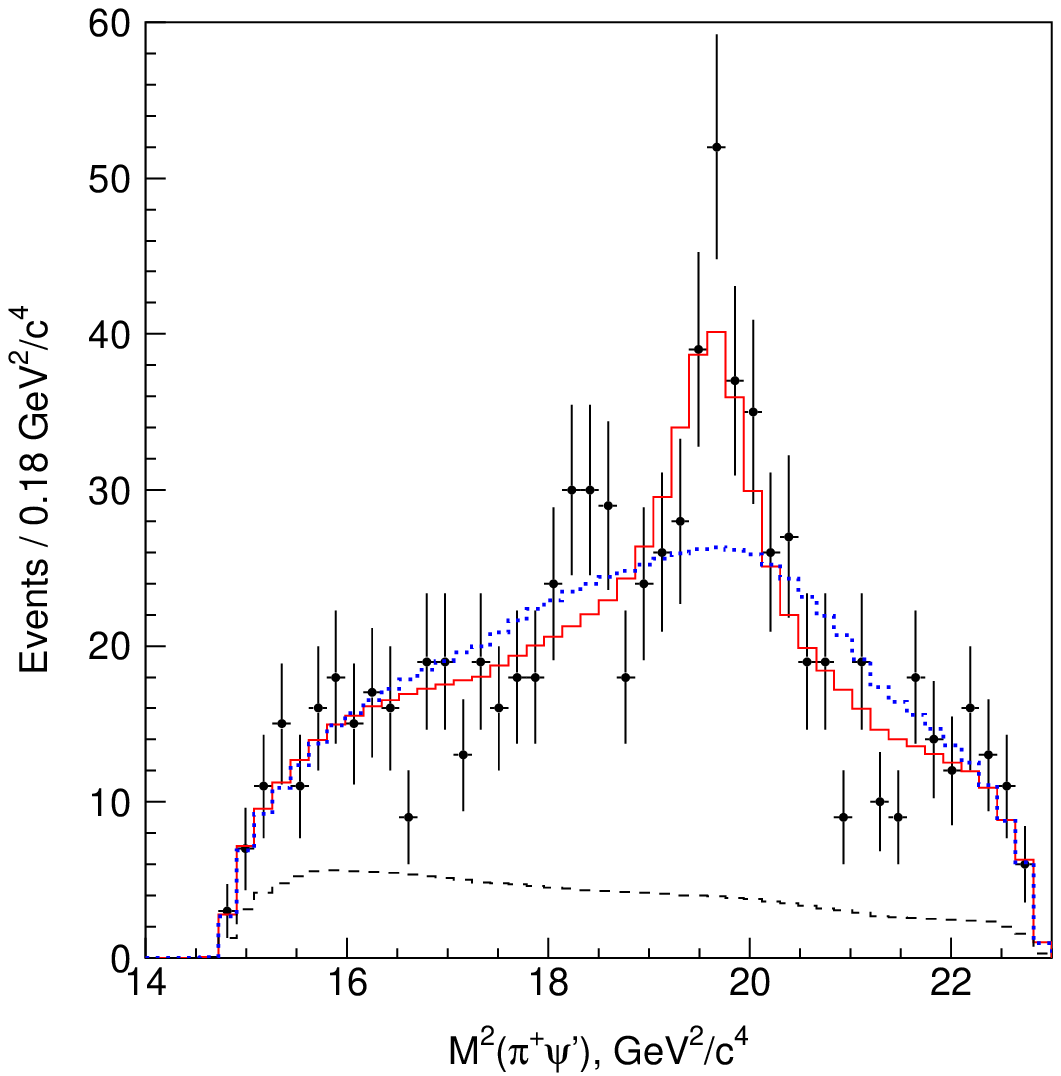}
\caption{The data points show the $M^2(\pi\psip)$ projection of the Dalitz
plot with the $K^*$ bands removed.  The histograms show the corresponding
projections of the fits with and without a $Z\rt\pi\psip$ resonance term.}  
\label{fig:z4430_dalitz-analysis}
\end{figure}

\subsection{Two charged $Z$ peaks in the $\pi^+\chi_{c1}$ channel}

In addition to the $Z(4430)^+$, Belle has presented results of an analysis
of $B\rt K\pi^+\chi_{c1}$ decays that require two resonant states in the
$\pi^+\chi_{c1}$ channel~\cite{belle_z14050}.  
The $M^2(K\pi)$ {\it vs.} $M^2(\pi\chi_{c1})$ Dalitz plot,
shown in Fig.~\ref{fig:z4050_dalitz}, shows vertical bands of events
corresponding to $K^*(890)\rt K\pi$ and $K_2^*(1430)\rt K\pi$, plus a
broad horizontal band near  $M^2(\pi\chi_{c1})\simeq17.5$~GeV$^2$,
indicating a possible resonance in the $\pi^+\chi_{c1}$ channel.  In
this case, this horizontal band corresponds to 
$\cos\theta_{\pi}\simeq 0$, a location where interference between
partial waves in the $K\pi$ channel can produce a peak and, thus,
a detailed Dalitz analysis is essential.

\begin{figure}[htb]
\centering
\includegraphics*[width=60mm]{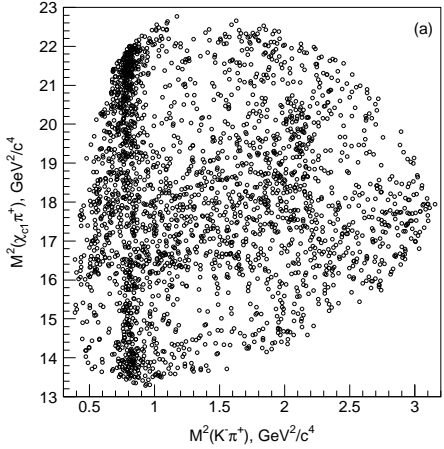}
\caption{The $M^2(K\pi)$ (horizontal) {\it vs.} $M^2(\pi\psi')$ (vertical) Dalitz plot distribution 
for candidate $B\rt K\pi\psi'$ events (from ref.~\cite{belle_z14050}).}
\label{fig:z4050_dalitz}
\end{figure}

In this case the kinematically allowed mass range for the $K\pi$ system extends beyond
the $K^*_3(1780)$ $F$-wave resonance and $S$-, $P$-, $D$- and $F$-wave terms for the $K\pi$
system are are included in the model.  The fit with a single resonance in the
$Z\rt \pi\chi_{c1}$ channel is favored over a fit with only $K^*$ resonances and no $Z$ by more than
$10\sigma$.  Moreover, a fit with two resonances in the $\pi\chi_{c1}$ channel is favored
over the fit with only one $Z$ resonance by $5.7\sigma$.  The fitted masses and widths of these
two resonances are: 
$M_1=4051\pm 14 ^{+20}_{-41}$~MeV and $\Gamma_1 = 82^{+21~~+47}_{-17~~-22}$~MeV and
$M_2=4248 ^{+44~+180}_{-29~~-35}$~MeV and $\Gamma_2 = 177^{+54~+316}_{-39~~-61}$~MeV.
The product branching fractions have central values similar to that for the $Z(4430)$ 
but with large errors.
Figure~\ref{fig:z4050_dalitz-analysis} shows the $M(\pi\chi_{c1})$ projection
of the Dalitz plot with the $K^*$ bands excluded and the results of the fit with no
$Z\rt\pi\chi_{c1}$ resonances and with two $Z\rt\pi\chi_{c1}$ resonances.

\begin{figure}[htb]
\centering
\includegraphics*[width=60mm]{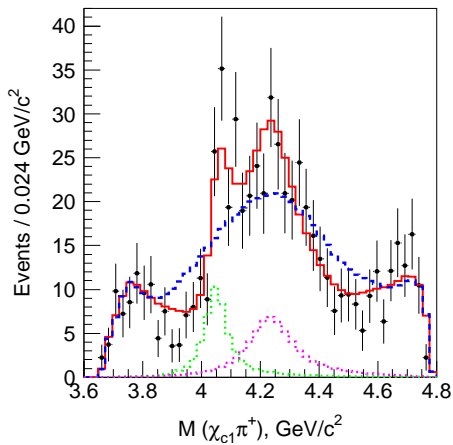}
\caption{The data points show the $M(\pi\chi_{c1})$ projection of the Dalitz
plot with the $K^*$ bands removed.  The histograms show the corresponding
projections of the fits with and without the two $Z\rt\pi\chi_{c1}$ resonance terms.}  
\label{fig:z4050_dalitz-analysis}
\end{figure}

\section{Summary}

The number of $XYZ$ states continues to grow.  Here I have reported on a new Belle
$X(3915)\rt\omega\jp$ mass peak in $\gamma\gamma\rt\omega\jp$~\cite{belle_y3915}.
In another talk at this meeting, Kai Yi reported 
on the CDF group's evidence for the $Y(4140)$, a narrow $\phi\jp$ resonance in
$B^+\rt K^+\phi\jp$ decays with mass $4143.0\pm 2.9\pm 1.2$~MeV and width
$11.7^{+8.3}_{-5}\pm 3.6$~MeV~\cite{CDF_y4140}.  The statistical significance for
this state is $3.8\sigma$ and it needs to be confirmed in other experiments.
However, this may not occur soon since, as Yi pointed out, the $B$-factory
experiments have poor acceptance for $B\rt K\phi\jp$, with with $\phi\jp$ in
this mass range.  

The mass and width of Belle's new $\omega\jp$ peak agrees well with BaBar's mass
and width values for the $``Y(3940)''\rt\omega\pi$ resonance seen in $B\rt K\omega\jp$
decays.  It is likely that these are the same state, and maybe we should
start calling the $``Y(3940)''$ the $Y(3915)$.  The lower mass would make this state 
more amenable to an assignment as the $\chi^{\prime}_{c0}$ charmonium state,  
but the large $\Gamma(Y\rt\omega\jp)$ partial width remains problematic. 
Belle expects to present an analysis
of $B\rt K\omega\jp$ decays with their full data sample ({\it i.e.} with nearly
four times the data that were used for the original $Y(3940)$ paper) sometime in 
the near future.  

As measurements of the masses of the $X(3872)$ and the $D^0$ meson improve, the 
$X(3872)$ gets closer and closer to the $m_{D^0}+m_{D^{*0}}$ mass threshold:
$M_{X(3872)}-m_{D^0}-m_{D^{*0}}= -0.35\pm 0.41$~MeV.  Braaten points out that
if the $J^{PC}$ of the $X(3872)$ is $1^{++}$, as seems most likely, this
nearness to the mass threshold implies that the $X(3872)$ has to be an $S$-wave
$D^0\bar{D^{*0}}$ molecular state with a huge, $\sim 6$~fermi, rms
separation.   It not clear how the production characteristics in high energy 
$p\bar{p}$ collisions of such a fragile extended object
could be so similar to those of the very
compact and tightly bound $\psip$ charmonium state.  Another question is 
how do the $c$ and $\bar{c}$ quarks in such widely separated
open charm mesons ever get close enough to form the $\jp$ that is produced
in the relatively frequent $\pipi\jp$ decay channel?

New data from Belle show no sign for any of the $1^{--}$ $Y$ states
decaying to $D^{**}\bar{D}$ final states, as would be expected if they
are $c\bar{c}$-gluon hybrid states.   In general, the total
lack of any sign of any signals for any of the $1^{--}$ $Y$ states in the 
$ D^{(*)}\bar{D}^{(*)}$ and $D\bar{D}^{(*)}\pi$ channels suggests
that the $\pipi\jp$ ($\pipi\psi'$) partial widths might well be much larger than the 
508~keV lower limit for the $Y(4260)$ presented in ref.~\cite{moxh}.
Any model that addresses these states should include
some mechanism to enhance the partial widths for these transition to vector charmonium states.  
Such a mechanism is not 
obviously present for $1^{--}$ $c\bar{c}$-gluon hybrids: for these, Lattice 
QCD calculations indicate that the $c\bar{c}$ pair is primarily in a 
spin-singlet state~\cite{dudek}.  Thus, rather than being enhanced,
transitions to a $\jpsi$ or  $\psi '$ are expected to be 
suppressed because of the required spin-flip of one of the charmed quarks.

If the charged $Z$ states reported by Belle in the $\pi^+\psip$ and $\pi^+\chi_{c1}$
channels are in fact meson resonances, they would be ``smoking guns'' for exotics.  It is
therefore important that the Belle results get confirmed by other experiments.  
BaBar made an extensive study of $B\rt K\pi^+\psip$ that neither
confirmed nor contradicted the Belle $Z(4430)^+$ result.  A similar BaBar study of
$B\rt K\pi^+\chi_{c1}$ might prove more conclusive.  CDF can access the $Z(4430)^+$
and we look forward to results from them in the near future.  In
the meantime, Belle remains confident that their analyses are sound and the
peaks that are seen in the $\pi^+\psip$ and $\pi^+\chi_{c1}$ invariant
mass distributions are not due to reflections from dynamics in the $K\pi$ system.

\section{A few final comments}

A number of theoretical models have been proposed for the $XYZ$ states: 
\begin{itemize}
\item  molecules, either of two open charmed mesons or of light mesons with charmonium;
\item  diquark-diantiquarks;
\item  $c\bar{c}$-gluon hybrids;
\item  hadroncharmonium, bound states of charmonium with highly excited light mesons.
\end{itemize}
\subsection{molecules}
Its closeness to the $D^{*0}\bar{D^0}$ mass threshold plus the
accumulating evidence for a $1^{++}$ $J^{PC}$ assignment make the identification
of the $X(3872)$ as a loosely bound $S$-wave $D^{*0}\bar{D^0}$ molecule 
inescapable~\cite{braaten_priv}.  Although some of the other states are near 
two-body thresholds ({\it e.g.} the $Y(4660)$ is near the $f_0(980)\psip$
threshold and has been attributed to an $f_0(980)\psip$ molecule~\cite{hanhart}),
this is not a universal feature of these states. One difficulty with interpreting a 
state as bound light meson plus charmonium system is the identification of a
binding mechanism.  The $\pi$, $\rho$,
$\omega$, etc. mesons do not couple to charmonium states and, thus,
normal nuclear-physics-like binding mechanisms do not apply.  

In a talk presented at this meeting, Raquel Molina presented an interesting model
that identified the $Y(3940$, $Z(3940)$ \& $X(4160)$ as dynamically generated
states produced by $D^*\bar{D^*}$ and $D_s^*\bar{D_s^*}$ interactions~\cite{molina}.
This model reproduces the measured masses of these states quite well, but does not
address other properties, like the large $\omega\jp$ partial width of the $Y(3940)$. 
In his talk,
Daniel Gamermann presented a dynamical model that forms the $X(3872)$ from $D^{*}\bar{D}$
interactions and explicitly addresses the decays of the $X(3872)$ to $\pipi\jp$
and $\pipi\pi^0\jp$~\cite{gamermann}.

\subsection{diquark-diantiquarks}
The diquark-diantiquark picture necessarily implies the existence of a rich array
of Isospin  and Flavor-$SU(3)$ partners for each of the $XYZ$ states.  To date,
no such partner states have been observed.

\subsection{$c\bar{c}$-gluon hybrids}
Problems with $c\bar{c}$-gluon hybrid assignments are discussed above.  Although
these continue to be the favored interpretation for the $1^{--}$ $Y$ states, this
is not because of any of their specific properties (other than their masses) that
have been measured to date.  $c\bar{c}$-gluon hybrids are necessarily electrically
neutral, so this interpretation does not apply to the charged $Z$ states.

\subsection{Hadrocharmonium}  
Dubynskiy and Voloshin have investigated a QCD
version of a van der Waal's force and found that it can be sufficiently strong
to bind light hadrons to a charmonium core in the case where the light hadron
is a highly excited resonance~\cite{voloshin}.  The resulting ``hadro-charmonium''
states would rather naturally have large partial widths for decays to light hadrons plus 
charmonium, which is a common feature of the $XYZ$ states.  However, this idea
has not been used to make any detailed predictions, so it is difficult to
evaluate its applicability.  Note that this scheme probably cannot be invoked to bind 
an $f_0(980)$ to a $\psip$ to form a $Y(4660)$ according to the suggestion
of ref.~\cite{hanhart} mentioned above, since the $f_0(980)$, a ground-state
scalar meson, is hardly a highly excited resonance.

\paragraph{~}
The $XYZ$ states remain a mystery and, therefore, continue to be interesting.

\section{Acknowledgments}
I thank Klaus Peters and the other organizers for arranging such an informative
meeting.  I also thank my Belle collaborators Ruslan Chistov, Galina Pakhlova,
Sadaharu Uehara and Changzheng Yuan for their help in the preparation of
my talk and this write-up.  This work has been supported in part by the
WCU program (R32-2008-000-10155-0) of the National Research Foundation of Korea.

%\begin{thebibliography}{9}   % Use for  1-9  references


\begin{thebibliography}{99}
\bibitem{conj} The inclusion of charge-conjugate modes is always implied.
Also, when two errors are presented, the first one is always statistical
and the second systematic.
\bibitem{mass-diff}
This is $2(m_d-m_u)/\cos 2\theta$, where 
$m_d$ ($m_u$) is the $d$-quark ($u$-quark) mass and  $\theta\simeq 20^0$
is a mixing angle.
\bibitem{theta-pi_def}  
$\theta_{\pi}$ is the angle between the $\pi^+$ and the negative of the $\psi'$ direction 
in the $K\pi$ rest frame.
\bibitem{babar-signif}
BaBar reports a $1.9\sigma$ signal with
mass and width similar to Belle's; if the mass and width are fixed at the Belle values,
the significance increases to $3.1\sigma$.
\bibitem{model_variations}
Variations in fit values from these alternative models are 
the main sources of the large systematic errors.
\end{thebibliography}
\end{document}